\documentclass{moriond}

\def\Journal#1#2#3#4{{#1} {\bf #2}, #3 (#4)}

\def\PLB{{\em Phys. Lett.}  B}
\def\PRL{\em Phys. Rev. Lett.}
\def\PRD{{\em Phys. Rev.} D}

\def\be{\begin{equation}}
	\def\ee{\end{equation}}
\def\bea{\begin{eqnarray}}
	\def\eea{\end{eqnarray}}

\usepackage{amsfonts}

\usepackage{bm}
\usepackage{hyperref}
\usepackage{mathrsfs}
\usepackage{graphicx}
\usepackage{empheq}
\usepackage{ulem}

\usepackage[usenames]{color}
\usepackage{ytableau}

\DeclareSymbolFontAlphabet{\mathrsfs}{rsfs}
\DeclareMathAlphabet{\mathcal}{OMS}{cmsy}{m}{n}

\newcommand{\nn}{\nonumber}
\newcommand\calO{\mathcal{O}}
\newcommand{\dd}{\mathrm{d}}

\newcommand{\tr}{\mathrm{Tr}}

\newcommand{\dtc}[1]{{\Bigl(\frac{\dd{#1}}{\dd t}\Bigr)}^{2}}

\newcommand{\Pial}[2]{\mathop{\Pi}_{#1}{}_{\!{#2}}}

\newcommand{\xiau}[2]{\mathop{\xi}_{#1}{}^{\!{#2}}}
\newcommand{\xiperpau}[2]{\mathop{\xi}_{#1}{}^{\!{#2}}_{\!\perp}}

\newcommand{\xidotperpau}[2]{\mathop{\dot{\xi}}_{#1}{}^{\!{#2}}_{\!\perp}}

\newcommand{\xiddotperpau}[2]{\mathop{\ddot{\xi}}_{#1}{}^{\!{#2}}_{\!\perp}}

\newcommand{\Eal}[2]{\mathop{E}_{#1}{}_{\!{#2}}}

\newcommand{\etaa}[1]{\mathop{\eta}_{#1}}

\newcommand{\vau}[2]{\mathop{v}_{#1}{}^{\!{#2}}}

\newcommand{\ta}[1]{\mathop{t}_{#1}}

\newcommand{\bmxiperpa}[1]{\mathop{\bm{\xi}}_{#1}{}_{\!\perp}}

\newcommand{\Jau}[2]{\mathop{J}_{#1}{}^{\!{#2}}}

\newcommand{\rhostara}[1]{\mathop{\rho}_{#1}{}_{\!\!*}}
\newcommand{\omegaa}[1]{\mathop{\omega}_{#1}}

\newcommand{\bse}{\begin{subequations}}
	\newcommand{\ese}{\end{subequations}}

\newcommand{\Photo}{}

\begin{document}
	\vspace*{4cm}
	\title{Dipolar dark matter theory based on a non-Abelian Yang-Mills field}
	
	\author{Luc Blanchet$^{a}$, Emeric Seraille$^{a,b}$}

	\address{
		$^{a)}$ 
		${\mathcal{G}}{\mathbb{R}}\varepsilon{\mathbb{C}}{\mathcal{O}}$, Institut d'Astrophysique de Paris, UMR 7095, CNRS, 
		\\98\textsuperscript{bis} boulevard Arago, 75014 Paris, France.\\
		$^{b)}$ Laboratoire de Physique de l’Ecole Normale Sup{\'e}rieure, ENS, CNRS, Universit{\'e} \\ PSL, Sorbonne Universit{\'e}, Universit{\'e} Paris Cit{\'e}, F-75005 Paris, France}
	
	\maketitle\abstracts{Most theories that attempt to reproduce the Modified Newtonian Dynamics (MOND) phenomenology for dark matter at galactic scales rely on \textit{ad hoc} free functions, preventing them from being regarded as fundamental. In this work, we present a new theory that reproduces MOND, built on a supposed to be fundamental Yang-Mills gauge field based on SU(2), with a gravitational coupling constant, and emerging in a low-acceleration regime, below the MOND acceleration scale. The gauge field plays the role of the internal force in the dipolar dark matter (DDM) model. We discuss how certain solutions of this theory recover the deep MOND regime without introducing arbitrary functions in the action. Within this framework, the MOND phenomenology appears to be due to the existence of a new sector of particle physics.}

	\section{Introduction}
	
	The MOND phenomenology, first introduced by Milgrom in 1983,~\cite{Milg1} could be an alternative way to understand dark matter, as it impressively describes the observed dynamics of most (spiral and elliptical) galaxies. This includes the flat rotation curves of galaxies, the baryonic Tully-Fisher relation for spirals, and the tight correlation between the presence of dark matter and the level of acceleration below the MOND acceleration scale $a_{0} \approx 1.2 \times 10^{-10}\,\text{m\,s}^{-2}$. These observations remain mostly unexplained by the standard $\Lambda$CDM cosmological model. The MOND phenomenology is summarized by Milgrom's law, where a function $\mu$ interpolates between the strong acceleration regime where dark matter effects are irrelevant and a weak acceleration regime (the so-called deep MOND regime) where the dynamics is modified according to
	\begin{equation}
		\label{Poisson_milgrom}	\bm{\nabla}\cdot\left[\mu\Bigl(\frac{g}{a_0}\Bigr) \bm{g} \right] = - 4 \pi G \rho_\text{bar}\,,
	\end{equation}
	with $\bm{g}=\bm{\nabla}U$ and $\rho_\text{bar}$ the baryon's mass density. However, despite the remarkable success at the galactic scale, MOND faces important and unexplained difficulties when extrapolating to the smaller solar system scale, and at the larger galaxy cluster scale.
	Significant efforts have been made to interpret Milgrom's law as the result of a modified gravity theory with additional fields and no dark matter.~\cite{Sand97,Bek04,SZ21,BS24} Nevertheless, it seems hopeless to consider these models as motivated by fundamental physics because they rely on free functions to recover the MOND formula~\eqref{Poisson_milgrom}. In this work, based on Ref.~\cite{BlanchetSeraille25}, we propose a different approach and introduce non-standard dark matter particles, interacting through an internal interaction in the form of a Yang-Mills gauge field, and inspired by the remarkable dielectric analogy of MOND.

	\section{The MOND dielectric analogy}\label{MondDielectricAnalogy}
	
	The MOND equation~\eqref{Poisson_milgrom} is formally equivalent to Gauss's law describing a dielectric medium, namely $\bm{\nabla}\cdot\bm{D}=\rho_\text{free}$, where $\rho_\text{free}$ is the density of free charges, and the electric displacement field $\bm{D}$ differs from the applied external field $\bm{E}$ due to the polarization of the medium: $\bm{D}=\epsilon_0\bm{E}+\bm{\Pi}_e$, where $\bm{\Pi}_e$ is the polarization related to the bound charges with $\rho_\text{bound} = - \bm{\nabla}\cdot\bm{\Pi}_e$. Introducing the electric susceptibility coefficient $\chi_e$ such that $\bm{\Pi}_e=\epsilon_0\chi_e\bm{E}$, we can write by analogy the MOND interpolating function as $\mu=1+\chi$, where $\chi(g)$ is interpreted as a coefficient of gravitational susceptibility. The gravitational polarization then reads
	\begin{equation}
		\label{Polarization}
		\bm{\Pi} = - \frac{\chi}{4\pi G} \,\bm{g}\,, 
	\end{equation}
	while the bound charges in electrostatics can naturally be interpreted as dark matter with $\rho_\text{DM} = - \bm{\nabla}\cdot\bm{\Pi}$. 
	To push this analogy a step further, a microscopic description of a DDM medium based on gravitational dipoles was proposed in Ref.~\cite{B07mond}. In particular, while in the electric case, the electric susceptibility is positive, corresponding to the screening of electric fields by polarization charges, in the gravitational case we should have $\chi<0$. Consequently, there is an anti-screening of ordinary masses by polarization masses, and therefore an enhancement of the gravitational field produced by the ordinary matter, which leads to a dark matter effect consistent with MOND. An unconventional aspect of the model is that the gravitational dipoles should be made of particles with positive and negative (gravitational-type) masses, and we have to consider doublets of sub-particles, one with positive gravitational mass $m_\text{g} = m$ and one with negative gravitational mass $m_\text{g} = - m$, while $m_\text{i} = m$ represents their always-positive inertial mass. The ordinary particle ($m_{g}>0$) is still attracted by ordinary matter, when the exotic particle ($m_{g}<0$) is repelled by ordinary masses. Furthermore, as the dipole constituents repel each other gravitationally, we need to invoke an attractive additional dark (non-gravitational) interaction acting only on dark matter constituents. 
	For this purpose we consider a toy model in which the particles interact with the Newtonian gravitational potential $U$ and with an internal potential $\Phi$.\footnote{We introduce a parameter $\eta$ to distinguish gravitational mass and internal charge.} The non-relativistic Lagrangian reads
	\begin{align}\label{toymodel0}
		L_\text{toy} &= \!\sum_{(m, m)} \!m \,\biggl[ \frac{1}{2}\bm{v}^{2}+U(\bm{y})+ \eta  \Phi(\bm{y}) \biggr] + \!\sum_{(m, -m)} \!m \,\biggl[ \frac{1}{2}\overline{\bm{v}}^{2}-U(\overline{\bm{y}})- \eta \Phi(\overline{\bm{y}}) \biggr] 
		\,,
	\end{align}
	where the sums run over the doublets of particles $(m_\text{i}, m_\text{g}) = (m, \pm m)$, and where $\bm{v}=\dd\bm{y}/\dd t$ and $\overline{\bm{v}}=\dd\overline{\bm{y}}/\dd t$. The interactions between these particles create microscopic dipole moments $\bm{\xi} = \bm{y} - \overline{\bm{y}}$ defined as the separation between the two types of particles; the position of the dipolar particle is the center of inertial mass, $\bm{x}=\frac{1}{2}(\bm{y} + \overline{\bm{y}})$. Expanding this Lagrangian to second order in the dipole moment, we obtain
	\begin{equation}\label{toymodel}
		L_\text{toy}=\sum_\text{dipoles} m \biggl[ \bm{v}^{2} + \frac{1}{4}\dtc{\bm{\xi}} + \bm{\xi}\cdot\bm{\nabla}\bigl(U + \eta \Phi\bigr) + \mathcal{O}(\xi^{3})\biggr]\,.
	\end{equation}
	
	\section{A non-relativistic dark matter model}
	
	Now, we want to model the internal interaction specific to the dipole moments. We introduce a Yang-Mills (YM) gauge field associated with the gauge group SU(2), and the covariant derivative
	\begin{align}\label{covder}
		\mathcal{D}_\mu =  \nabla_\mu  + \frac{1}{\ell} K_\mu\,, \qquad  \text{with} \qquad 	K_\mu = \sum_a \mathop{K}_{a}{}_{\!\!\mu} \,\ta{a}\,,
	\end{align}
	where $\ell$ is a constant having the dimension of a length, and the YM field $K_{\mu}$ is an element of the SU(2) Lie algebra.\footnote{The generators of the algebra are defined such that $[t_{a},t_{b}]=- \sum_c \epsilon_{a b c} t_{c}$ (with $a,b,c=1,2,3$) and normalized  such that $\tr(t_a t_b)=-\frac{1}{2}\delta_{ab}$, where $\epsilon_{abc}$ is the totally antisymmetric Levi-Civita symbol with $\epsilon_{123}=1$.} The associated field strength tensor is then defined as
	\begin{align}
		H_{\mu\nu} \equiv \mathcal{D}_\mu K_\nu - \mathcal{D}_\nu K_\mu = \partial_\mu K_\nu - \partial_\nu K_\mu + \frac{1}{\ell}\bigl[K_\mu, K_\nu\bigr] = \ell\,\bigl[\mathcal{D}_\mu, \mathcal{D}_\nu\bigr]
		\,.
	\end{align}
	To describe the internal interaction and the coupling between this extra field and the dark matter, we adopt an Effective Field Theory (EFT) approach and consider all the terms preserving the SU(2) symmetry and parity invariance up to cubic order in the field strength tensor. We also assume that the YM field is a (non-Abelian version) of a graviphoton in the sense of Ref.~\cite{Scherk}, with a coupling constant given by the gravitational constant G, thus
	\begin{align}\label{LYM}
		L_\text{YM} = \tr\biggl\{ c^2 \hat{\Pi}^{\mu\nu} H_{\mu\nu} + \frac{c^4}{8\pi G} \biggl[ - \frac{\Lambda}{2} + H_{\mu\nu} H^{\mu\nu} + \alpha \,H_{\mu\tau} H^{\tau}_{\phantom{\tau}\nu}\!\stackrel{\star}{H}{}^{\!\!\mu\nu} + \calO(\alpha^2) \biggr]\biggr\}\,,
	\end{align}
	where the antisymmetric tensor $\hat{\Pi}_{a}^{\mu\nu} = \frac{\eta_{a}}{c}(\Jau{a}{\mu}\xiau{a}{\nu} - \Jau{a}{\nu}\xiau{a}{\mu})$ can be interpreted as a polarization tensor, with $J_{a}^{\mu}$ the conserved current associated with the dipole, and $H^{\star\mu\nu}=\varepsilon^{\mu\nu\rho\sigma}H_{\rho\sigma}$.  
	We have also introduced an EFT length scale $\alpha$ associated with the expansion of the field strength tensor. The parameter $\Lambda$ plays the role of the cosmological constant and naturally scales as $\Lambda\sim 1/\alpha^2$. After having specified the internal sector, we need to add a model for the dark matter particles. We propose a generalization of the toy model~\eqref{toymodel}, considering three doublets of particles coupled with the internal YM field and the gravitational potential $U$. In the non-relativistic limit ($c \to +\infty$), the model describing all the relevant physics reads
	\begin{align}\label{LagN}
		L &= -\frac{1}{8\pi G} \,\partial_{i}U \partial_{i}U+ \rho_{\text{bar}}^{*}\Bigl(U + \frac{v_\text{bar}^2}{2}\Bigr) + \sum_a \rhostara{a} \Bigl( \vau{a}{2}+ \frac{1}{4} \xidotperpau{a}{i}\!\xidotperpau{a}{i} \Bigr) \nn\\& + \sum_a \biggl\{ \Pial{a}{i} \left( \partial_{i}U + \etaa{a}\Eal{a}{i}\right) + \frac{1}{8\pi G}\biggl[ \Eal{a}{i} \Eal{a}{i} +  \frac{\alpha}{2 c^2} \sum_{b,c}\epsilon_{abc} \,\epsilon_{ijk} \Eal{a}{i} \Eal{b}{j} \Eal{c}{k} \biggr]\biggr\} + \calO\left(\xi^3_\perp\right)\,.
	\end{align} 
	We recognize the Newtonian kinetic term and the standard interaction with the baryons. In the non-relativistic limit, the polarization tensor reduces to the ordinary polarization vector, say $\Pi_a^i=\rho^{*}_{a}\xi_{a\perp}^i$, and its coupling to $\partial_iU$ follows from~\eqref{toymodel}, with the dipole moment projected perpendicularly to the four-velocity of the particles, as indicated by $\perp$. The coupling between the polarization and the gravitational field $\partial_iU$ is again motivated by~\eqref{toymodel}. The parameter $\eta_a$ is the ratio of the coupling between the YM charge and the gravitational mass for each pair associated with the YM index $a$.
	
	Finally, in the non-relativistic approximation the YM field is described by the ``electric field'' $E_{a}^{i}$ such that the field strength tensor reduces to $H_{a}^{0i} = \frac{1}{c^2} E_{a}^{i} + \calO\left(c^{-4}\right)$, with the ``magnetic'' components $H_{a}^{ij}$ being negligible. Note that the last term in~\eqref{LagN}, cubic in the electric field, coming from the cubic term in~\eqref{LYM}, will \textit{in fine} be responsible for the MOND effect. 
	
	We draw particular attention to the fact that the model~\eqref{LagN} cannot be ``covariantized'' in standard general relativity, due to the two terms $\Pi_a^i \partial_{i}U$ and $\rho_a^*\,v_a^2$ in the dark sector, as for instance the term $\Pi_a^i \partial_{i}U$ can be eliminated by going to a frame freely falling with the baryons. Hence, the model implies a violation of the equivalence principle in the weak acceleration regime, which is a consequence of the negative gravitational masses introduced in the DDM sector.
	
	Varying~\eqref{LagN} with respect to the dipole moment and the YM electric field, we obtain
	\begin{align}
		\label{Eq_dipole_order2}
		\xiddotperpau{a}{\,i} + \omegaa{a}{}^{\!2} \xiperpau{a}{\,i} = 2 g_i - \frac{3 \bar{\alpha}}{8} \sum_{b,c} \epsilon_{abc} \frac{\eta_a}{\eta_b\eta_c}\omegaa{b}{}^{\!2}\!\omegaa{c}{}^{\!2}\epsilon_{ijk}\xiperpau{b}{\,j} \xiperpau{c}{\,k} + \calO\left(\xi^3_\perp\right)\,, 
	\end{align}
	where we have introduced the analogue of the ``plasma'' frequency of the DDM medium $\omega_{a}^{2} = 8\pi G \eta_{a}^2 \rho^{*}_{a}$. Varying with respect to the gravitational potential $U$ we obtain a modified Poisson equation given by 
	\begin{align}\label{ModifiedPoissonD}
		\bm{\nabla}\cdot\bm{\mathcal{D}} = - 4\pi G \rho^{*}_{\text{bar}}\,, ~\quad \text{where} \quad  \bm{\mathcal{D}} = \bm{\nabla}U - 4\pi G \sum_a \rho^{*}_{a} \bmxiperpa{a} \, .
	\end{align}
	
	\section{Generic solutions reproducing MOND}
	
	In this last section, we show that we can find generic solutions that reproduce the MOND phenomenology in the deep MOND regime. For this purpose, we introduce simplifying ansatz :
	\begin{enumerate}
		\item  
		The three YM mass ratios of the DDM particles satisfy $\sum_{a} \frac{1}{\eta_{a}^2} \simeq 1$ .
		
		\item We assume that one of the particles (labeled by the YM index 1, by convention) plays a privileged role, such that there is a hierarchy between the plasma frequencies:
		\begin{equation}\label{assumeobs}
			{(\omegaa{2}-\omegaa{3})}^2~\ll~ \omega_\text{obs}^2 ~\ll~ {\omegaa{1}}^2 ~\ll~ {(\omegaa{2}+\omegaa{3})}^2\,,
		\end{equation}
		where $\omega_\text{obs}=2 \pi/T_\text{obs}$ is associated with the observation time range of the dark matter effects at the scale of galaxies.\footnote{This hierarchical assumption is not as strong as we might expect, as the equations of motion associated with the dark matter imply that the acceleration of the particles is only sensitive to the tidal gravitational field. Thus, in a cosmological evolution scenario, we expect that the particle densities stay, to first approximation, at their cosmological value, allowing, for instance $\omega_{2}-\omega_{3}$ to remain small, and the hierarchy to remain satisfied.}
	\end{enumerate}
	For a given galaxy, we may restrict the problem to be locally unidimensional, with the gravitational field $g=U'(x)$ in the direction of the $x$-coordinate, pointing toward the center of the galaxy. In this case, the modified Poisson equation~\eqref{ModifiedPoissonD} becomes
	\begin{align}\label{deepMOND}
		\bigl({\mathcal{D}}^x\bigr)' = \Bigl(-\frac{3\bar{\alpha}}{8}\frac{k}{\eta_1\eta_2\eta_3} \,g^2\Bigr)' =  - 4\pi G \rho_{*\text{bar}} \,,
	\end{align}
	where $k$ is a dimensionless parameter associated with the initial conditions at the beginning of the formation of the galaxy. The equation~\eqref{deepMOND} represents exactly the deep-MOND limit of the MOND equation, with an effective acceleration constant  
	\begin{align}\label{expressiona0}
		a_0 = - \frac{8}{3\bar{\alpha}}\frac{\eta_1\eta_2\eta_3}{k} \,.
	\end{align}
	The MOND acceleration we obtain is \textit{a priori} non-universal, as it may depend on the formation scenario of the galaxy, and could be different, for instance, between spiral and elliptical galaxies. More work (probably numerical) should be done to assess the ``universality'' of the MOND scale~\eqref{expressiona0}. We also obtain that $\Lambda \sim a_{0}^2/c^4$ naturally has the order of magnitude of the measured cosmological constant. Finally, we note that the theory presented here is a pure ``deep-MOND-limit'' theory, as it remains disconnected from the usual regime of Newtonian gravity or GR at high accelerations.

\end{document}